\documentclass[12pt]{article}
\pdfoutput=1
\usepackage{putex}

\usepackage{amsmath,amssymb,amsfonts}
\usepackage{amscd}
\usepackage{psfrag}
\usepackage[all,cmtip]{xy}
\usepackage{color}
\usepackage{mathtools}
\usepackage{amssymb}
\usepackage[square, comma, compress,numbers]{natbib}
\usepackage[]{amsmath}
\usepackage{graphicx}
\usepackage{slashed}
\usepackage[]{latexsym}
\usepackage{amscd}
\usepackage{mathrsfs}
\usepackage{mathrsfs}
\usepackage[margin=10pt,font=small,labelfont=bf]{caption}
\usepackage{simplewick}
\usepackage{bbold}
\usepackage{changepage}
\usepackage{multirow}

\numberwithin{equation}{section}

\newcommand{\be}{\begin{eqnarray}}
\newcommand{\ee}{\end{eqnarray}}
\newcommand{\bea}{\begin{eqnarray}}
\newcommand{\eea}{\end{eqnarray}}

\newcommand{\es}[2] {\begin{equation} \label{#1} \begin{split} #2 \end{split} \end{equation}}
\newcommand{\abs}[1]{\left\lvert #1 \right\rvert}

\def\Red{\color [rgb]{0.9,0.1,0.1}}

\def\Green{\color [rgb]{0.1,0.6,0.1}}

\def\beq{\begin{equation}} 
\def\eeq{\end{equation}} 
\def\<{\langle}
\def\>{\rangle}

\def\bZ {\mathbb{Z}} 
\def\cO {{\cal O}}

\def\cN {{\cal N}}

\newcommand{\ed}{\,.}
\newcommand{\ec}{\,,}

\allowdisplaybreaks

\usepackage{braket}
\usepackage{setspace}
\usepackage{array}
\usepackage{titlesec}

\titlespacing\section{0pt}{12pt plus 4pt minus 2pt}{6pt plus 2pt minus 2pt}
\titlespacing\subsection{0pt}{12pt plus 4pt minus 2pt}{6pt plus 2pt minus 2pt}
\titlespacing\subsubsection{0pt}{8pt plus 4pt minus 2pt}{4pt plus 2pt minus 2pt}

\newcolumntype{x}[1]{>{\centering\arraybackslash\hspace{0pt}}p{#1}}

\usepackage{lipsum}

\begin{document}

\preprint{PUPT-2491}

\institution{PU}{Joseph Henry Laboratories, Princeton University, Princeton, NJ 08544}

\title{Bootstrapping $O(N)$ Vector Models with Four Supercharges in $3\leq d \leq4$}

\authors{Shai M.~Chester, Luca V.~Iliesiu, Silviu S.~Pufu, and Ran Yacoby}

\abstract{We analyze the conformal bootstrap constraints in theories with four supercharges and a global $O(N)\times U(1)$ flavor symmetry in $3 \leq d \leq 4$ dimensions. In particular, we consider the 4-point function of $O(N)$-fundamental chiral operators $Z_i$ that have no chiral primary in the $O(N)$-singlet sector of their OPE\@.  We find features in our numerical bounds that nearly coincide with the theory of $N+1$ chiral super-fields with superpotential  $W = X \sum_{i=1}^N Z_i^2$, as well as general bounds on SCFTs where $\sum_{i=1}^N Z_i^2$ vanishes in the chiral ring.}
\date{}

\maketitle

\newpage

\tableofcontents 

\newpage
\section{Introduction and Summary}
\label{INTRO}

In the past few years there have been numerous numerical studies of the conformal bootstrap \cite{Polyakov:1974gs,Ferrara:1973yt,Mack:1975jr} based on the pioneering work \cite{Rattazzi:2008pe}. The results of these studies seem to suggest that even a small subset of the crossing symmetry constraints can be very restrictive when supplemented with a few conditions on the  conformal field theory (CFT) data. For example, the assumption that a 3d CFT has $\bZ_2$ or $O(N)$ global symmetry and only a small number of relevant operators is sufficient in order to determine many terms in the operator product expansion (OPE) of these operators to a great accuracy \cite{ElShowk:2012ht,El-Showk:2014dwa,Gliozzi:2014jsa,Kos:2014bka,Rattazzi:2010yc,Kos:2013tga,chester2015bootstrapping,Kos:2015mba}.\footnote{When the theory has a $\bZ_2$ global symmetry, the above assumptions define the 3d Ising universality class, and indeed the results of the numerical bootstrap are consistent with, and improve upon other numerical methods that rely on some microscopic definition of the Ising critical point. Similarly, when the CFT has $O(N)$ global symmetry, the assumptions define the universality class of the critical $O(N)$-vector model.}  The possibility that only a small subset of the constraints satisfied by a CFT may be needed to fix a large amount of CFT data deserves much more scrutiny. In particular, to assess whether this possibility is realized more generally, it would be useful to study the bootstrap in examples for which some precise information on the CFT data is already known. Supersymmetric CFTs (SCFTs) provide us with a plethora of such examples. Indeed, supersymmetry often allows one to determine exactly the dimensions of protected operators and some of their correlators. This information can be compared to the results of a conformal bootstrap analysis, or supplemented to it as an additional constraint, making SCFTs into excellent laboratories for assessing how restrictive the crossing symmetry and unitarity conditions are \cite{Alday:2013opa, Bashkirov:2013vya, Beem:2013qxa,Chester:2014fya,Chester:2014mea,Beem:2014zpa,Berkooz:2014yda, Alday:2014qfa, Beem:2015aoa, bobev2015bootstrapping, bobev2015bootstrapping2, chester2015accidental, poland2015exploring}.

In this paper we will focus on theories with four Poincar\'e supercharges, for which the dimensions of chiral operators and 2-point functions of conserved currents are calculable. For SCFTs with this amount of supersymmetry, the bootstrap constraints on the 4-point function of a chiral operator $X$ in general dimensions $2\leq d \leq 4$ were recently studied in \cite{Bobev:2015vsa,Bobev:2015jxa}. 

The work \cite{chester2015accidental} initiated a generalization of \cite{Bobev:2015vsa,Bobev:2015jxa} to SCFTs with a chiral superfield $Z_i$ in the fundamental irrep of an $O(N)$ global symmetry. The analysis \cite{chester2015accidental} had interesting implications for the theory containing both $X$ and $Z_i$, and with the most general $O(N)$ invariant cubic superpotential
\begin{align}
W=\frac{g_1}{2}X\sum_{i=1}^NZ_i^2 + \frac{g_2}6 X^3\,. \label{superpot}
\end{align}
In particular, it was shown that when $N>2$, the assumption of a supersymmetric RG fixed point of \eqref{superpot} with both $g_1 \neq 0$ and $g_2 \neq 0$ is inconsistent with unitarity and crossing symmetry in dimensions $3\leq d \leq 4$.\footnote{In 3d, if $N=2$ and $g_2=0$ then \eqref{superpot} is equivalent to the $XYZ$ model, in which $g_2$ is an irrelevant deformation. If $N=1$, \cite{chester2015accidental} argued that the theory flows to two decoupled copies of the 3d $\cN=2$ Ising SCFT.}  The likely explanation of this exclusion is that either $g_1$ or $g_2$ flows to zero in the IR\@. As a result, the IR SCFT has an enhanced symmetry.  The $4-\epsilon$ expansion suggests that it is $g_2$ that flows to zero in the IR, and that the $O(N) \times \mathbb{Z}_3$ flavor symmetry of \eqref{superpot} is enhanced to $O(N) \times U(1)$.  Under the $O(N)$ symmetry, the $Z_i$ form a fundamental vector and $X$ is a singlet, while  under the $U(1)$ symmetry the $Z_i$ carry charge $+1$ and $X$ carries charge $-2$.  

In this work, our goal is to use the bootstrap to pin-point the non-trivial SCFTs that are expected to arise in the IR limit of \eqref{superpot} with $g_1 = g \neq 0$ and $g_2=0$, 
 \es{Wtilde}{
  W = \frac{g}{2}X\sum_{i=1}^N Z_i Z_i \,.
 }  
To do this, we supplement the bootstrap constraints of the 4-point function $\langle Z_i\bar{Z}_jZ_k\bar{Z}_l\rangle$ studied in \cite{chester2015accidental} with CFT data that can be inferred for this specific model. Most importantly, the superpotential \eqref{Wtilde} implies the chiral ring relation $\sum_{i=1}^N Z_i^2\sim 0$, and that the $O(N)$-fundamental $Z_i$ is charged under the additional $U(1)$ flavor symmetry mentioned above. In addition, one can use supersymmetric localization and the results of  \cite{Closset:2012vg,Closset:2012ru} to calculate the ``central charges'' $c_T$, $c_J^{O(N)}$, and $c_J^{U(1)}$ defined as the ratio between the two-point functions of the canonically normalized stress tensor, $O(N)$ current, and $U(1)$ current in our SCFT and the corresponding quantities in a reference SCFT\@. 

The more notable results of our numerical analysis are as follows:

\begin{itemize}

  \item {\bf Bounds on the lowest-lying unprotected $O(N)$ singlet.} We provide upper bounds, as a function of $\Delta_{Z_i}$, on the dimension of the lowest-lying unprotected $O(N)$ singlet that appears in the $Z_i \times \bar Z_j$ OPE\@.  In the specific model \eqref{Wtilde}, this operator is a linear combination of $\abs{X}^2$ and $\sum_{i} \abs{Z_i}^2$ that at large $N$ has scaling dimension close to $2$ for any $d$.  Quite nicely, when we impose the values of the central charges corresponding to the IR SCFT fixed point of \eqref{Wtilde}, we find that for a given value of $N$ the allowed regions are spiked around the value of $\Delta_{Z_i}$ that can be computed exactly using $F$-maximization and supersymmetric localization.
  
A nice application of these bounds is to test the accuracy of various approximation schemes such as the $4-\epsilon$ expansion.  Indeed, the scaling dimension of the lowest-lying unprotected singlet has been calculated up to 4 loops in the $4-\epsilon$ expansion.  These results are hardly converged when setting $\epsilon = 1$, and various resummation techniques are in order.  As an example, we find certain Pad\'e resummations to be inaccurate as they give estimates for the scaling dimension that lie within the region disallowed by our bootstrap bounds.  Perhaps a better resummation technique or an expansion to higher orders in $\epsilon$ is required.  

	\item {\bf SCFTs with $\sum_{i=1}^N Z_i^2\sim 0$ in the chiral ring.} Imposing the chiral ring relation $\sum_i Z_i^2 \sim 0$ that follows from \eqref{Wtilde} leads to universal lower bounds on $\Delta_{Z_i}$.  For instance, for $N=1$ our results imply that in any 3d $\cN=2$ SCFT with a chiral operator $Z$ such that $Z^2\sim 0$ in the chiral ring, we have that to a good approximation $\Delta_{Z}\geq \frac{2}{3}$. This bound is saturated for the Ising-SCFT studied in \cite{Bobev:2015vsa,Bobev:2015jxa}, and we conjecture that it is exact. It would very interesting to prove this conjecture in the future. In \cite{Bobev:2015vsa,Bobev:2015jxa}, the chiral ring relation  was also found to be satisfied at a kink whose position varies smoothly within $2\leq d\leq 4$; the interpretation of this kink in $d>2$ is currently not known.\footnote{In $d=2$ this mysterious kink merges with another kink whose position smoothly interpolates to the 3d $\cN=2$ Ising SCFT. Exactly at $d=2$ its position coincides with the $\cN=(2,2)$ minimal model with $c=1$.} In 4d it corresponds to a feature in the $\cN=1$ bootstrap bounds that was originally discovered in \cite{Poland:2011ey}, and recently explored further in \cite{poland2015exploring}. We find a family of similar kinks in $3\leq d \leq 4$ for every value of $N\leq N_\text{crit.}(d)$ that we checked, where $N_\text{crit.}(3)=2$, $\lim_{d\to 4}N_\text{crit.}(d)\to\infty$, and in general $N_\text{crit.}(d)$ increases with $d$. In $d=4$ we extrapolate the position of these kinks to $N \rightarrow \infty$ where they approach $\Delta_{Z_i} \approx 1.2$. This value may guide the search for an explicit possible realization of these features.

	\item {\bf Bounds on $c_J^{O(N)}$ and a generalization of $F$-maximization.}  We also present numerical lower bounds on the $c_J^{O(N)}$ central charge in terms of the scaling dimension $\Delta_{Z_i}$ of the chiral operator $Z_i$.  In our previous work in \cite{chester2015accidental}, we presented such bounds for $O(N)$-invariant SCFTs in $d=3$ when no other theory-specific information was assumed.  Here, we generalize the results of~\cite{chester2015accidental} by performing a similar analysis in non-integer $d$ and imposing more theory-specific information such as the values of the various other central charges.  
	
	It is worth commenting on the extension to non-integer $d$.  In the expressions for the (super)conformal blocks, $d$ appears as a parameter, so it is possible to extend these equations to non-integer $d$, as was done in a supersymmetric context in \cite{Bobev:2015jxa}.  A different, perhaps unrelated extension to non-integer $d$ was performed by Giombi and Klebanov \cite{Giombi:2014xxa} who provided an interpolation between the 3-sphere free energy $F$ in $d=3$ and the anomaly coefficient $a$ in $d=4$.  Assuming a similar procedure for computing various conserved current central charges as in integer dimensions \cite{Closset:2012vg}, one can further use the proposal of \cite{Giombi:2014xxa} to compute these quantities in non-integer $d$ \cite{chester2015accidental}.  Interestingly, we find that the values of $c_J^{O(N)}$ computed using this procedure almost saturate our numerical bounds determined using the extension of the superconformal blocks to non-integer $d$ of \cite{Bobev:2015jxa}.

\end{itemize}

The rest of this paper is organized as follows.  In Section~\ref{THEORY}, we review localization and $\epsilon$-expansion results for our models, as well as the setup for our numerical bootstrap analysis. In Section~\ref{numerics}, we present numerical bounds for various values of $N$ on the 2-point function coefficients of the $O(N)$, $U(1)$, and stress-tensor conserved currents, as well as on the scaling dimensions of the chiral operator $Z_i$ and the unprotected lowest dimension non-trivial $O(N)$-singlet scalar operator in the $Z_i\times\bar{Z}_j$ OPE.

\section{The Model}
\label{THEORY}

Consider an $\cN=2$ supersymmetric theory in 3d containing the $N+1$ chiral superfields $X$ and $Z_i$ ($i=1,\ldots,N$), and with a superpotential \eqref{Wtilde}.   
The model defined by \eqref{Wtilde} is expected to flow in the IR to an interacting SCFT with an $O(N)\times U(1)$ flavor symmetry and an R-symmetry $U(1)_{R}$, whose charges are specified in Table \ref{charges}. In this section we will review some of the properties of the CFT data of the model \eqref{Wtilde} that will be used in the numerical analysis of Section \ref{numerics}.

\begin{table}[htp]
\begin{center}
  \begin{tabular}{c||c|c||c}
        & $O(N)$       & $U(1)$ & $U(1)_{R}$ \\
        \hline
  $Z_i$ & $\mathbf{N}$ &  $1$   &  $r_Z$ \\
  $X$   & $\mathbf{1}$ &  $-2$  &  $2-2r_Z$      
  \end{tabular}
  \end{center}
  \caption{Charges of fields in the model \eqref{Wtilde}.\label{charges}}
\end{table}

\subsection{Localization}

Let us start by reviewing what CFT data can be generally extracted for 3d $\cN=2$ theories when using supersymmetric localization. The R-symmetry charges determine the conformal dimensions of operators in the chiral ring of the SCFT. To determine these charges one has to identify the superconformal R-symmetry in the IR, which can be done using the procedure of $F$-maximization \cite{Jafferis:2010un}. Moreover, as described in \cite{Closset:2012vg,Closset:2012ru}, supersymmetry allows for the calculation of the coefficients of 2-point functions of conserved currents, which we refer to as `central charges'. The calculation of the R-charge $r_Z$ and of the flavor symmetry central charges use as an input the $S^3$ free energy $F=-\log |Z_{S^3}|$. On the other hand, the stress-tensor central charge can be determined through the partition function on the squashed sphere \cite{Closset:2012ru}. The partition functions on both manifolds are calculable using supersymmetric localization \cite{Kapustin:2009kz,Jafferis:2010un,Hama:2010av,Hama:2011ea}.

In this paper we will also be interested in studying the model \eqref{Wtilde} in dimensions $3\leq d\leq 4$. A formal extension of the superconformal algebra of theories with four Poincar\'e supercharges to $d$ dimensions was suggested in \cite{Bobev:2015jxa}. In this extension, chiral primaries can be defined in continuous $d$ and the relation of their dimension to their R-charge is given by
\begin{align}
\Delta = \frac{d-1}{2} r \ed \label{r2delta}
\end{align}
The authors of \cite{Bobev:2015jxa} also gave the dimensional continuation of superconformal blocks corresponding to 4-point functions of chiral operators, thus allowing for a generalization of the supersymmetric bootstrap to non-integer dimensions. In a different development, an extension of $F$ to the $S^d$ free energy in continuous $d$, denoted by $\tilde{F}$, was recently proposed in \cite{Giombi:2014xxa} for theories with only chiral superfields. Associated with $\tilde{F}$, the authors of \cite{Giombi:2014xxa} conjectured a generalized $\tilde{F}$-maximization procedure (as well as an $\tilde{F}$-theorem), which can be used to determine the R-charges of chiral operators in continuous $d$, and from \eqref{r2delta}, their scaling dimensions. As done in \cite{chester2015accidental}, we will also use $\tilde{F}$ to determine flavor symmetry central charges in non-integer dimensions.\footnote{It would be interesting to check the continuation of current 2-point functions as obtained through $\tilde{F}$, to those calculated by the more conventional $\epsilon$-expansion.} We cannot, however, determine the stress-tensor central charge in this way, since currently there is no known interpolation of the squashed sphere free energy in non-integer dimension.

Let us now list the CFT data discussed above in more detail for the particular model \eqref{Wtilde}. Let $j^\mu_{ij}$ and $j^\mu$ denote the $O(N)$ and $U(1)$ conserved currents, respectively, and $T^{\mu\nu}$ the stress-tensor. We define the central charges of these currents as
\es{currentsDef}{
  \langle j^\mu(x) j^\nu(0) \rangle &= c_J^{U(1)}\frac{\Gamma^2(d/2)}{4(d-1)(d-2)\pi^d} 
    I^{\mu\nu}(x)\frac{1}{x^{2d-2}} \,,\\
      \langle j^\mu_{ij}(x) j^\nu_{kl}(0) \rangle &= c_J^{O(N)}\frac{\Gamma^2(d/2)}{4(d-1)(d-2)\pi^d} \left(\delta_{ik}\delta_{jl}-\delta_{il}\delta_{jk}\right)
    I^{\mu\nu}(x)\frac{1}{x^{2d-2}} \,,\\
 \langle T^{\mu\nu}(x)T^{\rho\sigma}(0)\rangle&=c_T\frac{d\Gamma^2(d/2)}{4(d^2-1)\pi^d}\left(\frac{1}{2}\left(I^{\mu\rho}(x)I^{\nu\sigma}(x)+I^{\mu\sigma}(x)I^{\nu\rho}(x)\right)-\frac{1}{3}\eta^{\mu\nu}\eta^{\rho\sigma}\right)\frac{1}{x^{2d}}\,,
}
where $I^{\mu\nu}(x)= \eta^{\mu\nu}-2\frac{x^\mu x^\nu}{x^2}$. In our conventions, $c_J^{O(N)}= 1$ and $c_T=c_J^{U(1)}=N$ for a free $O(N)$-fundamental chiral multiplet of a single unit of $U(1)$ charge. 

The procedure of calculating $c_J^{O(N)}$ in $3\leq d\leq 4$ by using the $S^d$ free energy $\tilde{F}$ was described in \cite{chester2015accidental}, and can be trivially generalized to $c_J^{U(1)}$; we refer the reader to \cite{chester2015accidental} for more details. In Table \ref{val3} we list the values of $\Delta_{Z_i}$, $c_J^{O(N)}$ and $c_J^{U(1)}$ for some particular values of $N$ in dimensions $d=3$ and $d=3.5$, as well as values of $c_T$ in $d=3$.

\begin{table}[htp]

\begin{center}
$d=3$\\
\vspace{0.2cm}
  \begin{tabular}{c||c|c|c|c|c|c}
   $N$ & $1$ & $2$ & $3$ & $4$ & $10$ & $20$  \\
 \hline\hline
 $\Delta_{Z_i}$ &.708& .667 & .632 & .605  & .543 &.521\\
 \hline
 $c_J^{O(N)}$ & -- & .600 & .664 & .715  & .844 &.920\\
 \hline
 $c_J^{U(1)}$ &3.33 & 3.13 & 3.34 & 3.85  & 8.91& 18.63\\
 \hline
 $c_T$ &6.02& 8.72 & 11.85 & 15.31& 38.34&78.08\\ 
  \end{tabular}
  \end{center}

\begin{center}
$d=3.5$\\
\vspace{0.2cm}
  \hspace{-0.42cm}
   \begin{tabular}{c||c|c|c|c|c|c}
   $N$ & $1$ & $2$ & $3$ & $4$ & $10$ & $20$  \\
 \hline\hline
 $\Delta_{Z_i}$  &.851& .833 & .820 & .810 & .781&.767 \\
 \hline
 $c_J^{O(N)}$&  -- & .826 & .850 & .869 & .921 &.957\\
 \hline
 $c_J^{U(1)}$ & 4.27 & 4.76 & 5.41 & 6.15  & 11.48&21.18\\
  \end{tabular}
\end{center}
\caption{The scaling dimension of $\Delta_{Z_i}$, and the $O(N)\times U(1)$ flavor central charges at the infrared fixed point of \eqref{Wtilde} in $d=3, \, 3.5$.  The central charge of the stress-tensor $c_T$ is only  determined in $d=3$. 
The charges are normalized so that they equal $1$ and $N$, respectively, in a theory of $N$ free chiral multiplets. }
\label{val3}
\end{table}%

\subsection{$4-\epsilon$ Expansion}
\label{EPSILON}

In one of the numerical experiments we will perform in Section \ref{numerics} we place upper bounds on the dimension $\Delta'_{Ss,0}$ of the lowest dimension unprotected $O(N)$-singlet scalar in the $Z_i\times\bar{Z}_j$ OPE\@.  It is possible to obtain an approximation for the dimension of this operator using the $4-\epsilon$-expansion, as we now describe. 

Indeed, Wess-Zumino models with cubic superpotentials such as \eqref{Wtilde} have weakly coupled fixed points in $d=4-\epsilon$. These models were studied perturbatively in 4d in \cite{Ferreira:1996az,Ferreira:1997hx,Jack:1998iy,Jack:1999fa} up to 4-loop order, and their results can be adapted to our case of interest. For brevity, we will only present the results for our specific model to 3-loop order. In particular, the anomalous dimension matrix of $\bar{X}X$ and $\bar{Z}_iZ_i$ can be read off from the results of \cite{Ferreira:1996az,Jack:1998iy,Jack:1999fa}. One eigenvalue is always zero corresponding to the combination
\begin{align}
J_{U(1)} = 2\bar{X}X-\bar{Z}_iZ_i \ed
\end{align}
The operator $J_{U(1)}$ is the bottom component of the current multiplet corresponding to the flavor $U(1)$ symmetry, and is therefore not renormalized. To 3-loop order, the other eigenvalue turns out to be given by
\begin{align}
\gamma(g) = \frac{N+4}{16 \pi ^2}g^2 -   \frac{N+1}{32 \pi ^4} g^4 +  \frac{3 \left(N^2+N (6 \zeta (3)+11)+24 \zeta (3)+4\right)}{4096 \pi ^6}g^6 +O(g^8)\ed
\end{align}
Moreover, the 3-loop $\beta$-function of \eqref{Wtilde} in $d=4-\epsilon$ is 
\begin{align}
\beta_g = -\frac{\epsilon }{2} g +  \frac{ (N+4)}{32 \pi ^2} g^3 - \frac{N+1}{128 \pi ^4} g^5 +  \frac{N^2+6\zeta (3)(N+4) +11 N+4}{8192 \pi ^6} g^7 + O(g^9)\ed
\end{align}
Solving $\beta_g(g_*)=0$ we find the dimension of the unprotected scalar in the $\epsilon$-expansion to be
\es{eps}{
\Delta'_{Ss,0} = 2-\epsilon + \gamma(g_*) = 2 - \frac{4(N+1)}{(N+4)^2}\epsilon^2 + \frac{12(N+4)^2\zeta(3) + 2N (N(N-1)+16)}{(N+4)^2}\epsilon^3 + O(\epsilon^4)
}

For any quantity $f(d)$ known in the $\epsilon=4-d$ expansion up to a given order, we can construct a Pad\'e approximant
\es{pade}{
\text{Pad\'e}_{[m,n]}(\epsilon)=\frac{A_0+A_1\epsilon+A_2\epsilon^2+\dots+A_m\epsilon^m}{1+B_1\epsilon+B_2\epsilon^2+\dots+B_n\epsilon^n}\,,
}
where the coefficients $A_i,B_i$ are fixed by requiring that the expansion at small $\epsilon$ agrees with the known terms in $f(4-\epsilon)$. If a quantity is known in the $\epsilon$-expansion to order $k_0$, one can construct Pad\'e approximants of total order $m+n=k_0$.

In the $X^3$ superpotential case studied in \cite{Bobev:2015jxa}, the bootstrap results for the dimension of the lowest scalar in the $X\times\bar{X}$ OPE, agree to three digits with the $\text{Pad\'e}_{[2,1]}(\epsilon)$ or $\text{Pad\'e}_{[1,2]}(\epsilon)$ approximants obtained from the $\epsilon$-expansion evaluated to order $\epsilon^3$ \cite{AVDEEV1982321}. This suggests that a similar Pad\'e approximation could also be accurate in our case. Therefore, in the following sections 
we will use the $\text{Pad\'e}_{[2,1]}(\epsilon)$ approximant of \eqref{eps}.

\subsection{4-Point Function of $Z_i$ and Bootstrap}
\label{BOOTSTRAP}

Let us now describe the setup of our numerical analysis. Consider the 4-point function 
\begin{align}
\langle Z_i(x_1) \bar{Z}_j(x_2) Z_k(x_3) \bar{Z}_l(x_4) \rangle \ed \label{4pntTmp}
\end{align}
For the purpose of implementing the bootstrap we find it convenient to write $Z_i = Z_{1i}+iZ_{2i}$ and $\bar{Z}_i=Z_{1i}-iZ_{2i}$, and work with the real fields $Z_{Ii}$, treating $I=1,2$ as an $O(2)$ fundamental index.  The $O(2)$ symmetry here can be thought of either $U(1)_R$ or the flavor $U(1)$---the charges of $Z_i$ under both of these symmetries are proportional.   Instead of the 4-point function \eqref{4pntTmp} we can thus equivalently study 
\begin{align}
\langle Z_{Ii}(x_1) Z_{Jj}(x_2) Z_{Kk}(x_3) Z_{Ll}(x_4)\rangle \ed \label{4pt}
\end{align}

The crossing symmetry equations of \eqref{4pt} are identical to those appearing in Appendix~B of \cite{chester2015accidental}, to which the reader is referred for more details.\footnote{Note that the existence of an additional $U(1)$ symmetry compared to the model discussed in \cite{chester2015accidental} does not lead to additional crossing relations.} In compact form, the invariance of \eqref{4pt} under the exchange $(1,I,i)\leftrightarrow(3,K,k)$ implies the crossing equation
 \es{crossing}{
   0 &= \sum_{\cO\in Ss, \text{ all }\ell}  \lambda_\mathcal{O}^2 \vec V^{Ss}_{ \Delta, \ell}
      + \sum_{{\cO\in St, \text{ all }\ell }}  \lambda_\mathcal{O}^2 \vec V^{St}_{\Delta, \ell}
      + \sum_{{\cO\in Sa, \text{ all }\ell }}  \lambda_\mathcal{O}^2 \vec V^{Sa}_{ \Delta, \ell} \\
     &{}+  \sum_{\cO\in Ts, \text{ }\ell \text{ even }}  \lambda_\mathcal{O}^2 \vec V^{Ts}_{\Delta, \ell}
       + \sum_{\cO\in Tt, \text{ }\ell \text{ even }}     \lambda_\mathcal{O}^2 \vec V^{Tt}_{ \Delta, \ell}
       +  \sum_{\cO\in Ta, \text{ }\ell \text{ odd }}   \lambda_\mathcal{O}^2 \vec V^{Ta}_{\Delta, \ell} \,,
 }
where $\lambda_{\cO}$ are the OPE coefficients that must be real by unitarity, and $\vec V^{Rr}_{\Delta, \ell}$  are nine component vectors given by certain combinations of conformal blocks defined in \cite{chester2015accidental}. It is important that the $\vec V^{Rr}_{\Delta, \ell}$ also depend on $\Delta_{Z_i}$, though we will suppress this dependence in our notation to avoid clutter. The sums are over all superconformal multiplets in the $Z_{Ii}\times Z_{Jj}$ OPE, which are classified according to their $O(N)\times O(2)$ representation. The labels $S$ and $T$ denote, respectively, the singlet and rank-two traceless symmetric irreps of $O(2)$ (corresponding to operators that are uncharged or charged under $U(1)_R$ or the flavor $U(1)$, respectively). The singlet, rank-two traceless symmetric, and rank-two anti-symmetric irreps of $O(N)$ are denoted by $s$, $t$, and $a$, respectively.  Due to Bose symmetry, the operators in the irreps $Ts$ and $Tt$ must have even spins, those in $Ta$ should have odd spins, and those in all other irreps can have any integer spin.

The unitarity bounds in the different channels, as well as the additional constraints on the chiral ring imposed by the specific superpotential \eqref{Wtilde} are listed in Table~\ref{UnitarityTable}. These additional constraints are:
\begin{itemize}
\item The superpotential \eqref{Wtilde} implies that the chiral operator $\sum_{i=1}^N Z_i^2$ is a descendant. It is therefore removed from the $Ts$ sector (i.e., from the $Z_i\times\bar{Z}_j$ OPE).
\item In general, the $Z_i\times Z_j$ OPE can contain a dimension $d-2\Delta_Z$ conformal primary of the form $\cO\sim \bar{Q}^2\bar{\Psi}$, where $\bar{\Psi}$ is an anti-chiral primary of R-charge $r_{\bar{\Psi}} = 2r_Z-2$. In the models \eqref{superpot}, in the $Ts$ channel $\bar{Q}^2 \bar{\Psi}=\bar{Q}^2  \bar{X}$, while the $Tt$ channel does not contain a descendent $\bar{Q}^2 \bar{\Psi}$ of an anti-chiral primary.  Note that an anti-chiral primary $\bar \Psi$ of R-charge $2 r_Z - 2$ can satisfy the unitarity bound only if $\Delta_Z < d/4$.  In particular, in $d=4$ there is no such option if $Z_i$ are not free.
\end{itemize}

\begin{table}[htpb]
\centering
\begin{tabular}{c|c|c|c|}
\cline{2-4}
& \hspace{0.5in} $s$ \hspace{0.5in} & \hspace{0.5in} $t$ \hspace{0.5in}  & $a$  \\
\hline
\multicolumn{1}{|c|}{$S$} & \multicolumn{3}{c|}{$\Delta \geq \ell + d - 2$,\  \ for all allowed values of $\ell$} \\
\hline
\multicolumn{1}{|c|}{\multirow{4}{*}{$T$}} & \multicolumn{3}{c|}{$\Delta \geq \abs{2 \Delta_Z - (d-1) } + \ell + (d - 1)$, \ for all allowed values of $\ell$} \\
\multicolumn{1}{|c|}{} & \multicolumn{3}{c|}{$\Delta =2 \Delta_Z + \ell$, \ for all allowed values of $\ell$} \\
\cline{2-4}
\multicolumn{1}{|c|}{}  & \multicolumn{1}{c|}{$\Delta = d - 2 \Delta_Z$, \ for $\ell = 0$, $\Delta_Z \leq d/4$} &\multicolumn{1}{c|}{$\Delta = 2 \Delta_Z$, \ for $\ell = 0$} & \\
\hline
\end{tabular}
\caption{Unitarity bounds in different symmetry channels of the $Z_{Ii}\times Z_{Jj}$ OPE in the model \eqref{Wtilde}.}\label{UnitarityTable}
\end{table}

In our conventions, the relations between the OPE coefficients in \eqref{crossing} and the central charges defined in \eqref{currentsDef} are given by
\es{ctoOPE}{
c_J^{U(1)}=\frac{2^{2d-4}}{\lambda^2_{Ss,d-2,0}},\qquad c_J^{O(N)}=\frac{2^{2d-5}}{\lambda^2_{Sa,d-2,0}}, \qquad
c_T=\frac{2^{2d-1}}{(d-1)}\frac{\Delta_Z^2}{\lambda^2_{Ss,d-1,1}} \ec
}
where $\lambda_{Rr,\Delta,\ell}$ denotes the OPE coefficient of an operator of dimension $\Delta$ and spin $\ell$ transforming in the $Rr$ irrep of $O(N)\times O(2)$.  To find an upper bound on the OPE coefficient $\lambda_{R^*r^*,\Delta^*,\ell^*}$ of an operator $\cO^*$, we search for linear functionals $\vec{\alpha}$ satisfying the conditions
\es{lowerBound}{
&\vec{\alpha}(\vec{V}^{R^*r^*}_{\Delta^*,\ell^*})=1 \,, \\
&\vec{\alpha}(\vec{V}^{Rr}_{\Delta,\ell})\geq0\,\,  \text{for all $\vec{V}^{Rr}_{\Delta,\ell}\notin\{\vec{V}^{Ss}_{0,0},\vec{V}^{R^*r^*}_{\Delta^*,\ell^*}$\} \text{ and } $(\Delta,\ell)$ as in Table~\ref{UnitarityTable}}\,,
}
where $\vec{V}^{Ss}_{0,0}$ corresponds to the identity operator.
If such a functional $\vec{\alpha}$ exists, then along with the positivity of all $\lambda_{\cO}^2$ it implies that
 \es{UpperOPE}{
  \lambda_{R^*r^*,\Delta^*,\ell^*}^2 \leq - \alpha (\vec{V}^{Ss}_{0,0}) \ed
   }
To obtain the most stringent upper bound on $\lambda_{R^*r^*,\Delta^*,\ell^*}^2$, one should then minimize the RHS of \eqref{UpperOPE} under the constraints \eqref{lowerBound}.

To find upper bounds on the scaling dimension $\Delta'_{Ss, 0}$ of the lowest dimension non-identity $O(N)\times U(1)$-singlet scalar primary $\cO'_{Ss, 0}$ appearing in the $Z_{Ii} \times Z_{Jj}$ OPE, we consider linear functionals $\vec{\alpha}$ satisfying the following conditions:
\es{upperBound}{
&\vec{\alpha}(\vec{V}^{Ss}_{0,0})=1 \,, \\
&\vec{\alpha}(\vec{V}^{Rr}_{\Delta,\ell})\geq0\,\,  \text{for all $\vec{V}^{Rr}_{\Delta,\ell}\notin\{\vec{V}^{Ss}_{0,0},\vec{V}^{Ss}_{\Delta^*,0}\,|\, \Delta^* < \Delta'_{Ss,0}\}$ and $(\Delta,\ell)$ as in Table~\ref{UnitarityTable} }\,,\\
&\vec{\alpha}(\vec{V}^{Ss}_{d-2,0})\geq 0\,,
}
where the third condition allows for the existence of the conserved $U(1)$ flavor current of our model. The existence of any such $\vec{\alpha}$ would contradict \eqref{crossing}, and thereby allow us to find disallowed points in the $(\Delta_{Z_i},\Delta'_{Ss, 0})$ plane. If we set $\Delta'_{Ss, 0}$ to its unitarity value $d-2$, then we can find general bounds on $\Delta_{Z_i}$ as a function of $d$.

We can potentially strengthen the upper bounds on scaling dimensions of operators by inserting the known values of OPE coefficients $\{\lambda_{R_ir_i,\Delta_i,\ell_i}\}$ into the algorithm \eqref{upperBound}, so that we consider linear functionals $\vec{\alpha}$ satisfying the altered conditions:
\es{upperBound2}{
&\vec{\alpha}\left(\vec{V}^{Ss}_{0,0}+\sum_i\lambda^2_{R_ir_i,\Delta_i,\ell_i}\vec{V}^{R_ir_i}_{\Delta_i,\ell_i}\right)=1 \,, \\
&\vec{\alpha}(\vec{V}^{Rr}_{\Delta,\ell})\geq0\,\,  \text{for all $\vec{V}^{Rr}_{\Delta,\ell}\notin\{\vec{V}^{Ss}_{0,0}, \vec{V}^{R_ir_i}_{\Delta_i,\ell_i}, \vec{V}^{Ss}_{\Delta^*,0}\,|\, \Delta^* < \Delta'_{Ss,0}\}$ and $\Delta$ as in Table~\ref{UnitarityTable}}\,,\\
&\vec{\alpha}(\vec{V}^{Ss}_{d-2,0})\geq 0\,.
}
For instance, we can insert central charge values computed from localization using their relation \eqref{ctoOPE} to OPE coefficients.

\begin{figure}[p]
\begin{center}
 \includegraphics[width = 0.8\textwidth]{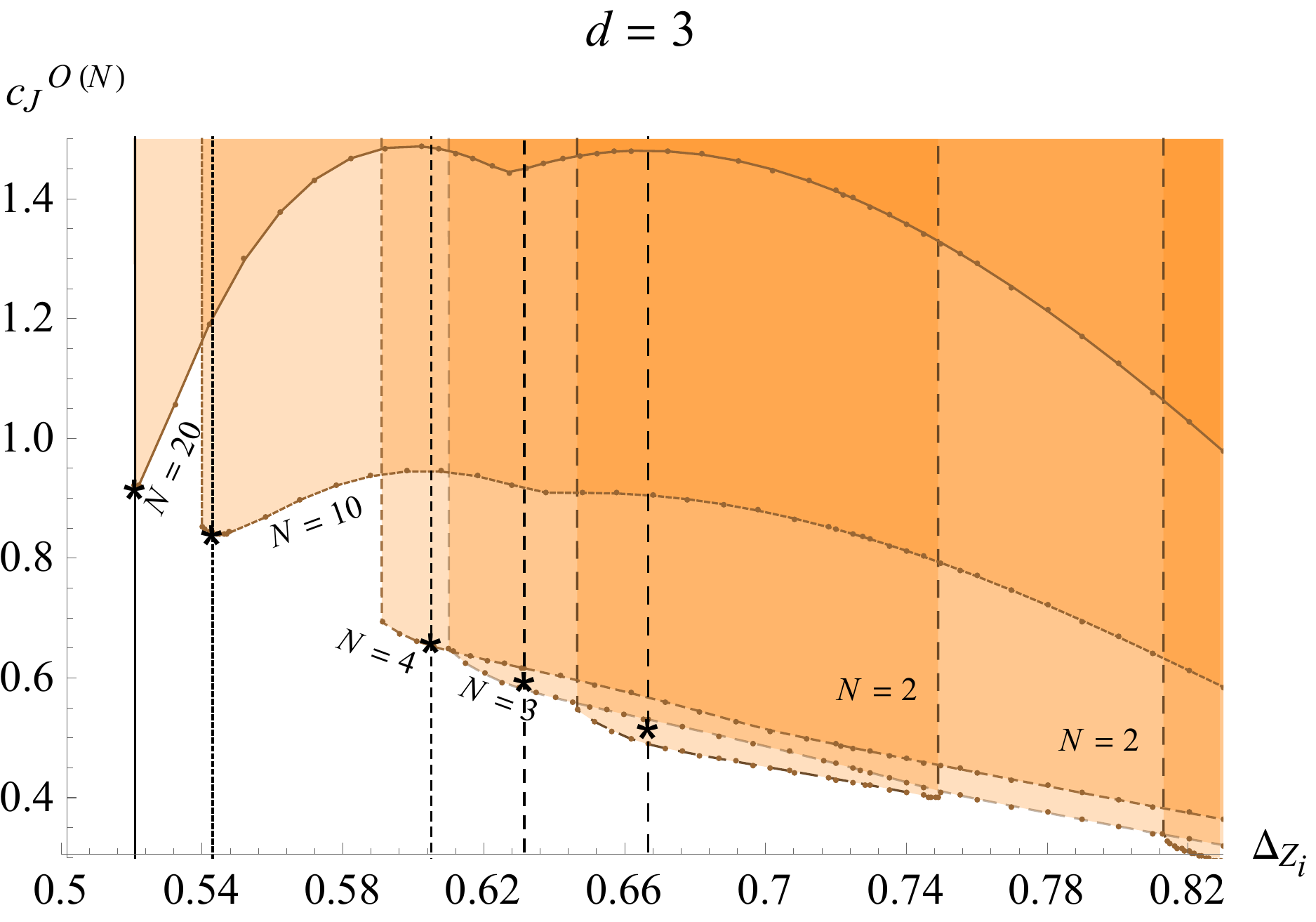} \includegraphics[width = 0.8\textwidth]{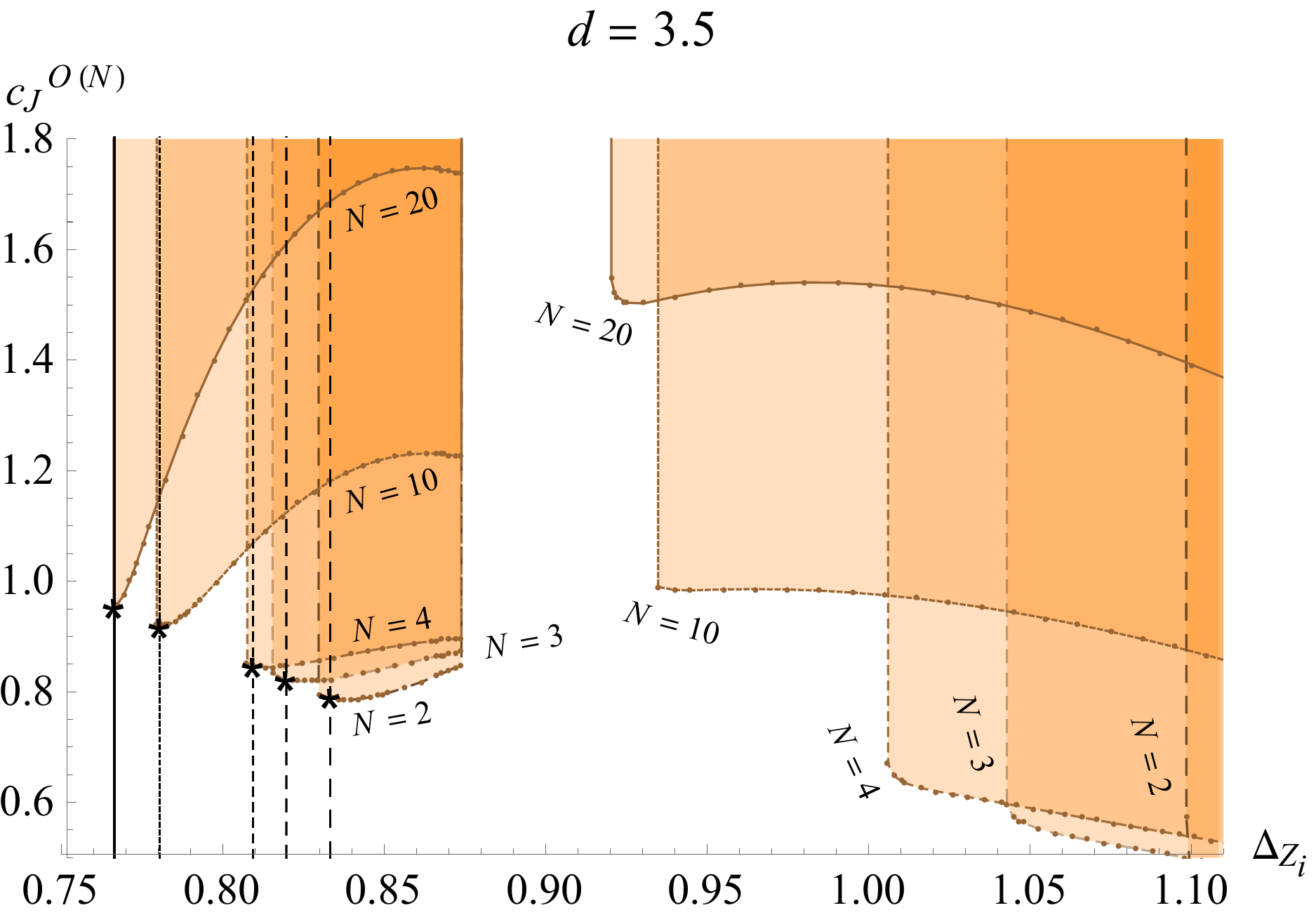}
 \caption{Lower bounds on the $O(N)$ flavor current central charge $c_J^{O(N)}$ as function of the scaling dimension $\Delta_{Z_i}$ of the chiral $O(N)$-fundamental primary in dimensions $d=3$ and $d=3.5$, for $N = 2, 3, 4, 10, 20$. Different shadings of orange denote the allowed regions for each $N$. The \textit{black} vertical lines denote localization values of $\Delta_{Z_i}$ for each $N$, while the asterisks denote the localization values of $c_J^{O(N)}$ for each $N$ (see Table \ref{val3}). These bounds were computed using $\ell_\text{max}=25$ and $\Lambda=19$. \label{fig:3dON}}
 \end{center}
 \end{figure}
 
\section{Numerical Bootstrap Results}
\label{numerics}

The problems \eqref{lowerBound}, \eqref{upperBound}, and \eqref{upperBound2} of finding functionals subject to inequalities can be rephrased as semi-definite programing problems as described in \cite{Poland:2011ey}, which we implemented using SDPB \cite{Simmons-Duffin:2015qma}.  In this section we will describe the results of our numerical analysis.

\subsection{Lower Bounds on central charges $c_{J}^{O(N)}$, $c_{J}^{U(1)}$ and $c_T$}
\label{chargeBounds}

\begin{figure}[p]
\begin{center}
 \includegraphics[width = 0.8\textwidth]{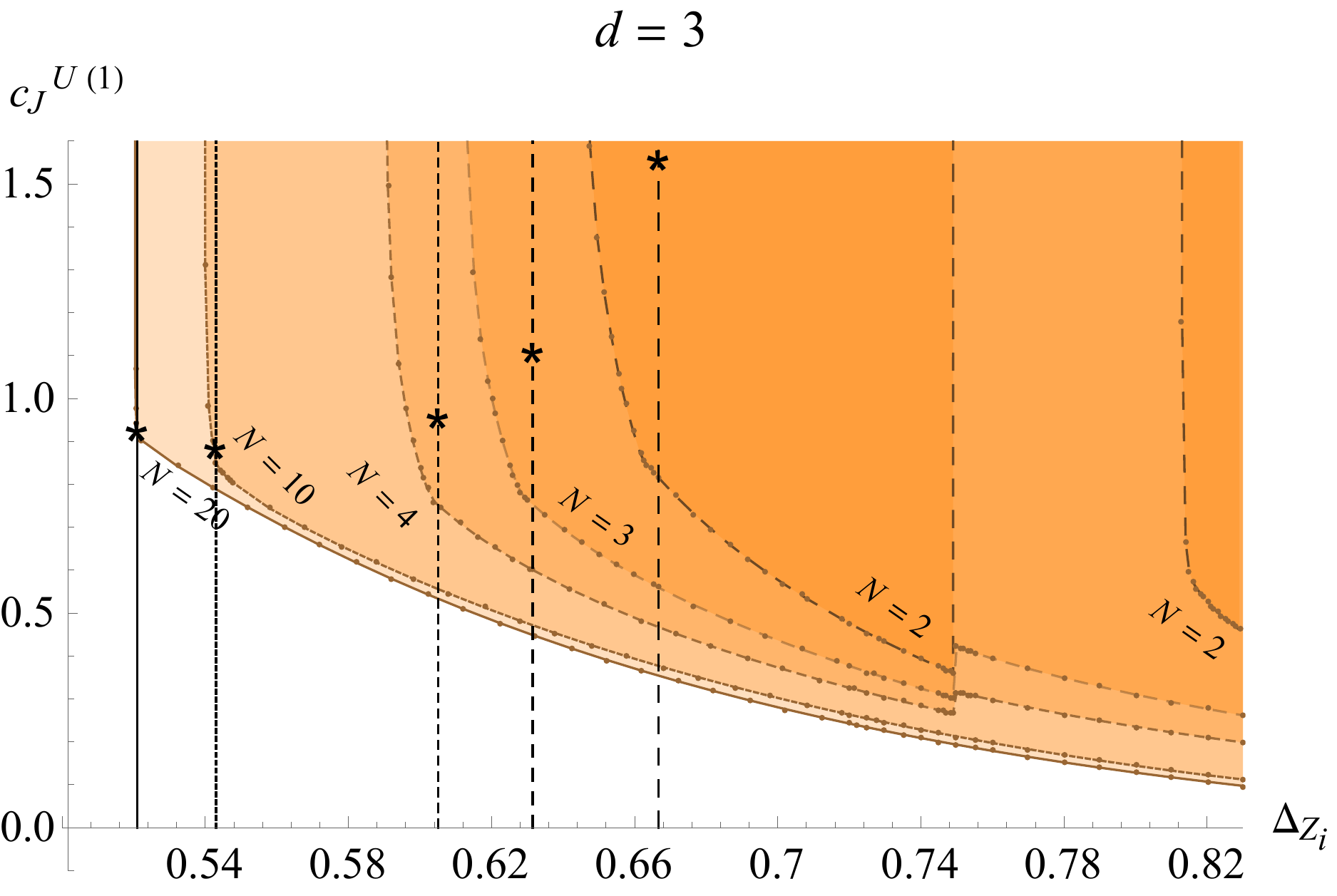}
  \includegraphics[width = 0.8\textwidth]{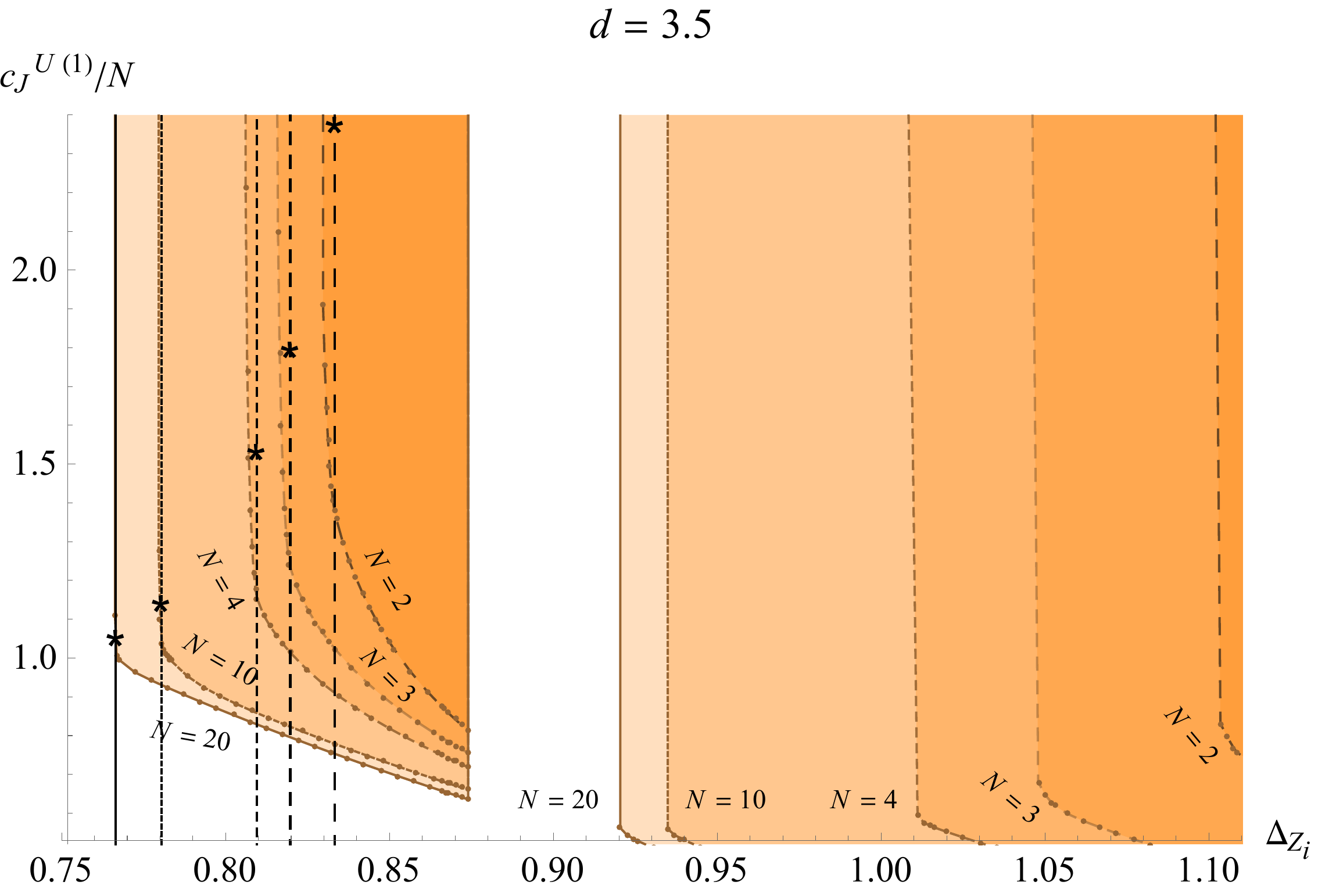}
 \caption{Lower bounds on the $U(1)$ central charge $c_J^{U(1)}$ in terms of the scaling dimension $\Delta_{Z_i}$ of the chiral $O(N)$-fundamental primary in dimensions $d=3$ and $d=3.5$, for $N = 2, 3, 4, 10, 20$. Different shadings of orange denote the allowed regions for each $N$. The \textit{black} vertical lines denote localization values of $\Delta_{Z_i}$ for each $N$, while the asterisks denote the localization values of $c_J^{U(1)}$ for each $N$ (see Table \ref{val3}). These bounds were computed using $\ell_\text{max}=25$ and $\Lambda=19$.    \label{fig:3dU1}}
 \end{center}
 \end{figure}
 
 \begin{figure}[p]
\begin{center}
 \includegraphics[width = 0.8\textwidth]{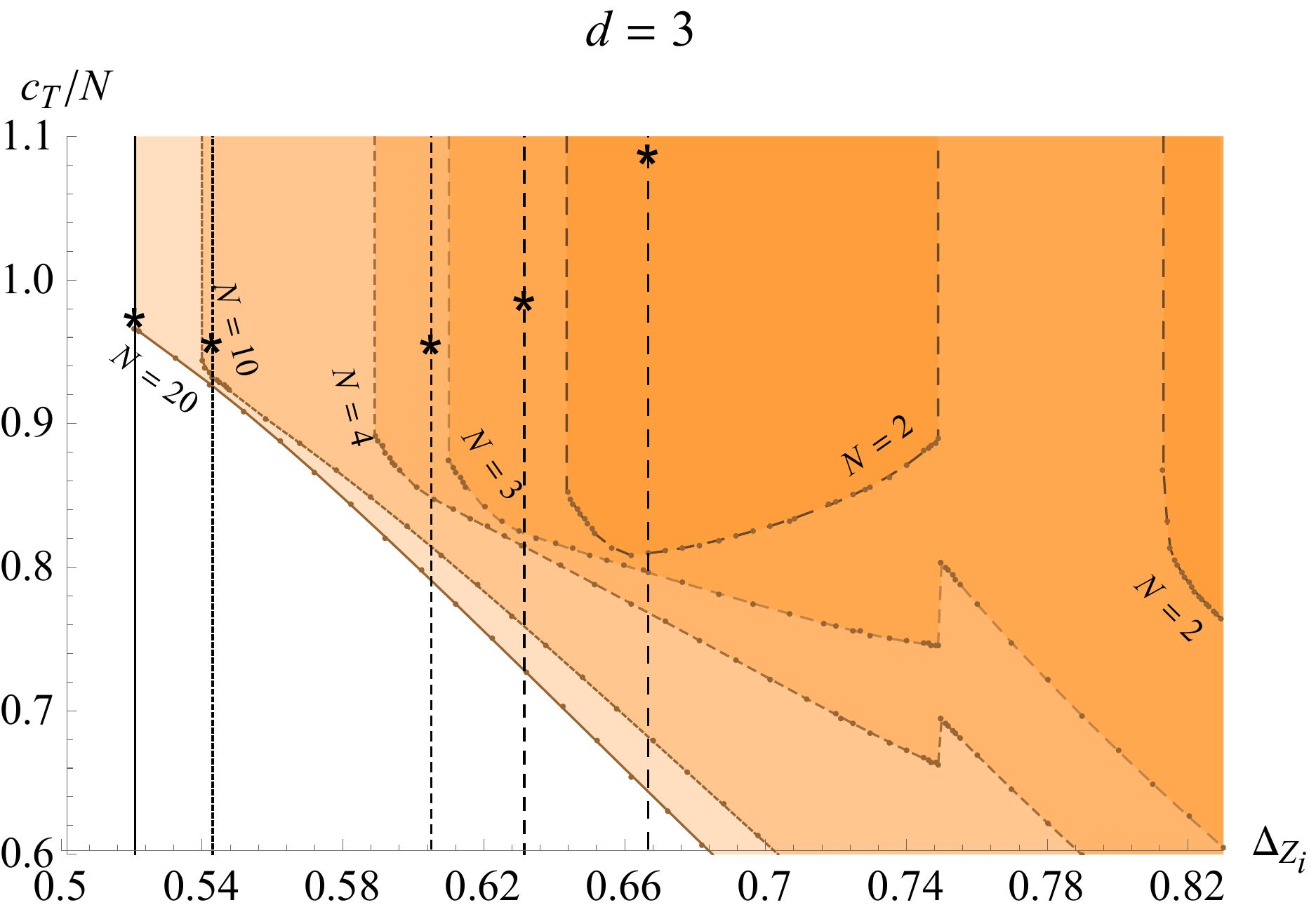}\\
  \includegraphics[width = 0.8\textwidth]{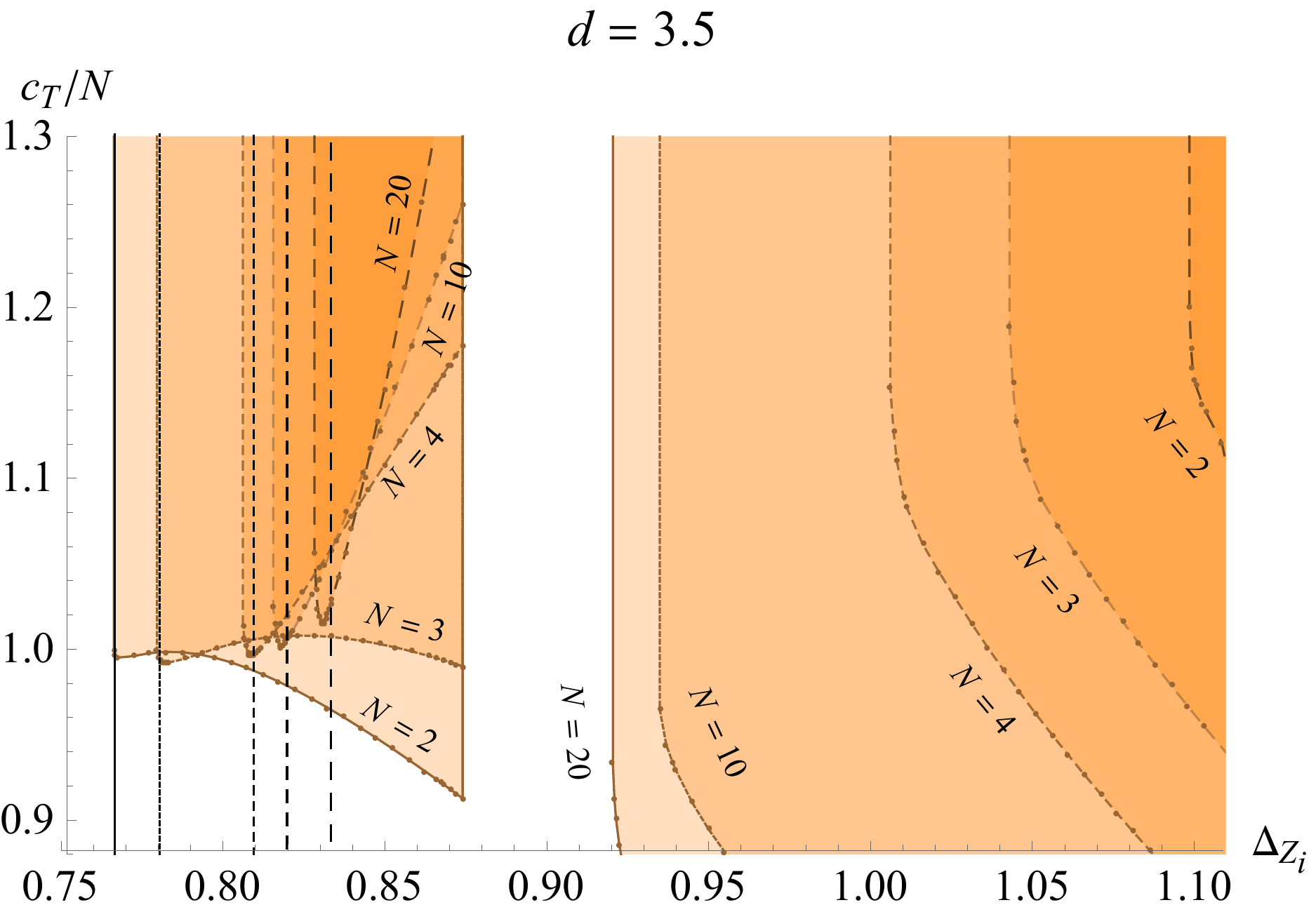}
 \caption{Lower bounds on central charge $c_T$ in terms of the scaling dimension $\Delta_{Z_i}$ of the chiral $O(N)$-fundamental primary in dimension $d=3$ and $d=3.5$, for $N = 2, 3, 4, 10, 20$. Different shadings of orange denote the allowed regions for each $N$. The \textit{black} vertical lines denote localization values of $\Delta_{Z_i}$ for each $N$, while the asterisks denote the localization values of $c_T$ for each $N$ (see Table \ref{val3}). These bounds were computed using $\ell_\text{max}=25$ and $\Lambda=19$.  \label{fig:3dcT}}
 \end{center}
 \label{3dcT}
 \end{figure}

\begin{figure}[t]
	 \begin{center}
	  \includegraphics[width = 0.85\textwidth ]{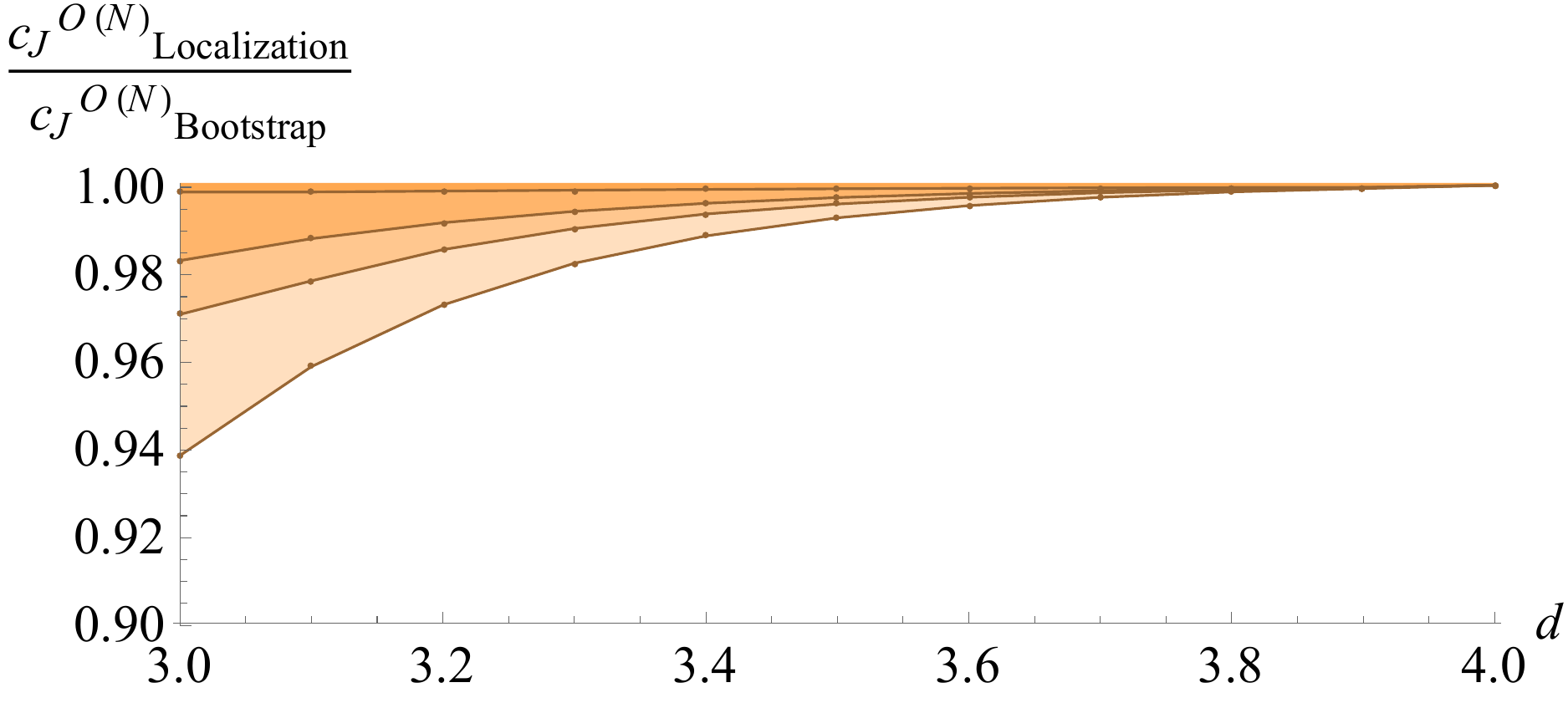}
	  \caption{\label{fig:dif} The ratio between the bootstrap lower bound on the coefficient $c_J^{O(N)}$ of the $O(N)$ current 2-point function and the value predicted using the generalization of the $F$-value of the 3d SCFT in the IR limit of \eqref{superpot} with $g_2=0$ to non-integer dimensions \cite{Giombi:2014xxa}. The curves, from bottom to top, correspond to $N = 2, \, 3, \, 4, \, 10$. 
	  }
	  \end{center}
	  \end{figure}

We start by presenting our bounds for the central charges $c_{J}^{O(N)}$, $c_{J}^{U(1)}$ and $c_T$ defined in \eqref{currentsDef}. In Figure \ref{fig:3dON} we show lower bounds on $c_J^{O(N)}$ as a function of $\Delta_{Z_i}$ in $d=3$ and $d=3.5$. For $d=3$ we see that the localization values of $c_J^{O(N)}$, given in Table \ref{val3}, nearly saturate these bounds, rapidly approaching kinks as we increase the value of $N$. For $d=3.5$, the values calculated using the generalization of $F$-maximization and supersymmetric localization to non-integer dimensions \cite{Giombi:2014xxa} appear on kinks for all the values of $N$ that we considered. In fact, as we show in Figure \ref{fig:dif}, these non-integer localization values nearly saturate the bootstrap bounds on $c_J^{O(N)}$ for the entire range $3\leq d\leq4$.

Figure \ref{fig:3dU1} depicts lower bounds on $c_J^{U(1)}$ as a function of $\Delta_{Z_i}$ in $d=3$ and $d=3.5$. For both $d=3$ and $d=3.5$, one can see kinks in the bounds for all the values of $N$ that we considered. However, compared with the $c_J^{O(N)}$ case (Figure \ref{fig:3dON}), the localization values of $c_J^{U(1)}$, given in Table \ref{val3}, saturate these bounds and approach the aforementioned kinks only for relatively high values of $N$. Similar bounds for $c_T$ are shown in Figure \ref{fig:3dcT}, in which the analytically determined values of $c_T$ also approach saturation relatively slowly as $N$ is increased. We conclude that the general bootstrap bounds on $c_J^{U(1)}$ and $c_T$ are not as optimal as those on $c_J^{O(N)}$ for the purpose of constraining the theories \eqref{Wtilde}. 

 A notable feature of all these central charge bounds is that in $d=3$ there is a gap in the allowed region for $N=2$ whenever $.75<\Delta_{Z_i}<.875$, while for the other values of $N$ the bounds continue across this range. In $d=3.5$ and for all values of $N$ studied, we find similar gaps in the allowed region starting at $\Delta_{Z_i}\approx.875$ and ending at different points depending on $N$. These gaps will be discussed further in Section \ref{chiralBounds}.

\subsection{Upper Bounds on Unprotected Scalar}
\label{singletBounds}

\begin{figure}[p]
\begin{center}
 \includegraphics[width = 0.8\textwidth]{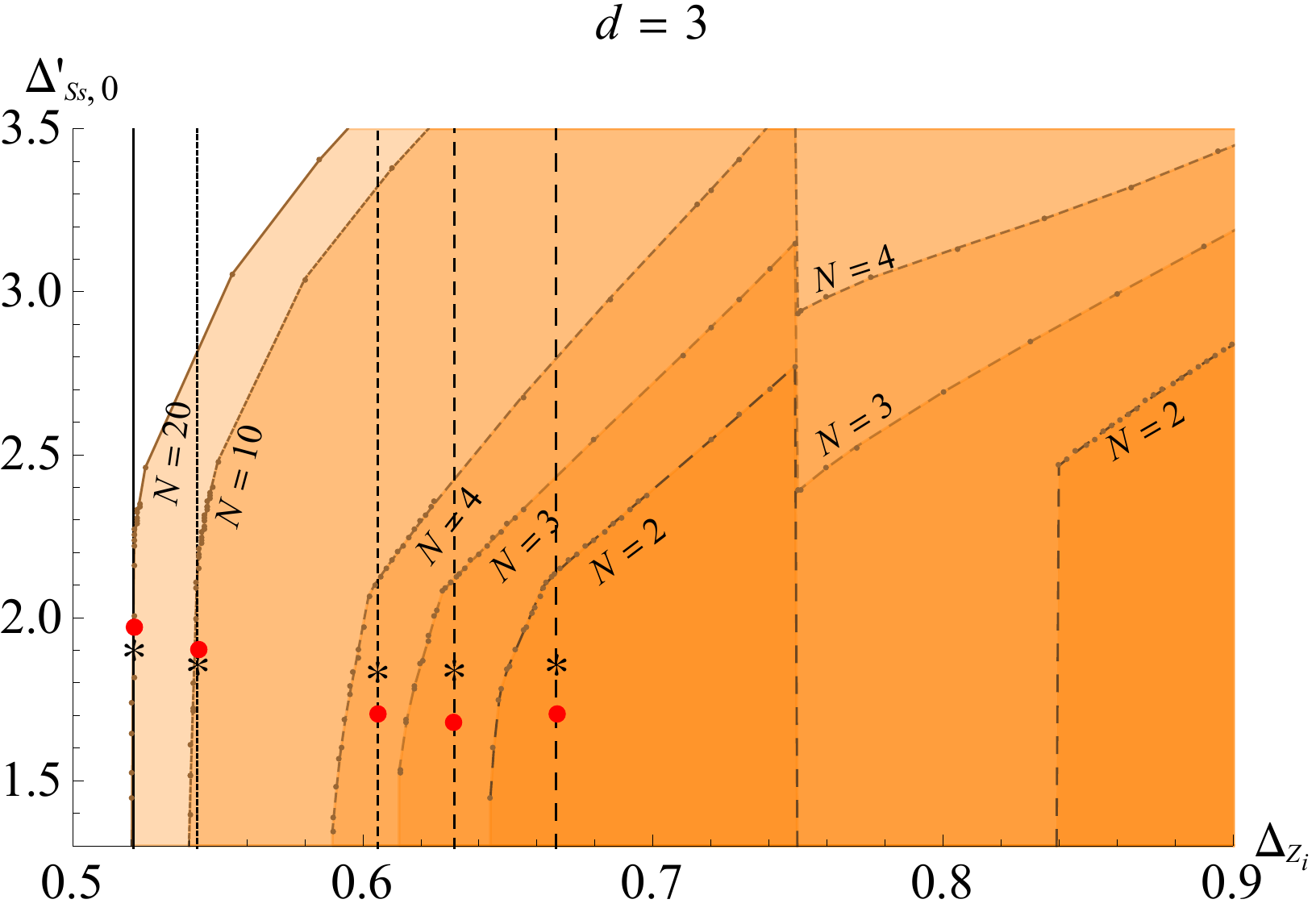}\\
  \includegraphics[width = 0.8\textwidth]{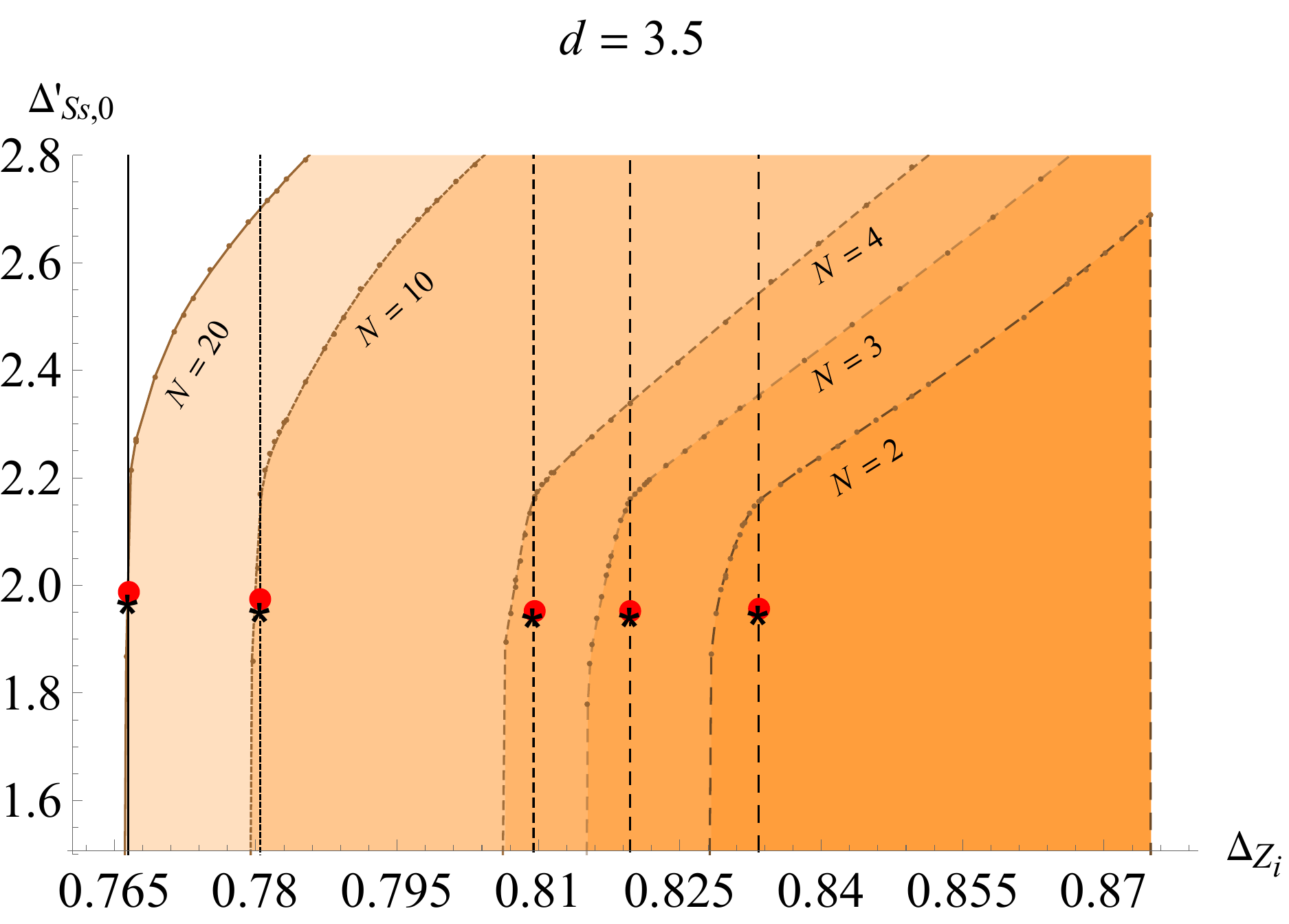}
 \caption{\label{fig:3dbin}Upper bound on the scaling dimension $\Delta'_{Ss, 0}$ of the lowest-lying $O(N)$-singlet scalar in the $Z_i\times\bar{Z}_j$ OPE, as a function of $\Delta_{Z_i}$ in dimensions $d=3$ and $d=3.5$, for $N = 2, 3, 4, 10, 20$. Different shadings of orange denote the allowed regions for each $N$. The \textit{black} vertical lines denote localization values (Table \ref{val3}) of $\Delta_{Z_i}$ for each $N$. The red dots denote the stronger bounds once the localization values of  $c_J^{U(1)} $ and $c_J^{O(N)}$ (Table \ref{val3}) are imposed on the spectrum. The asterisks indicate the Pad\'e approximation to the 3-loop $\epsilon$-expansion values of $\Delta'_{Ss, 0}$ for each $N$. Note that for $N=2$ there is a gap in the allowed region for $.75<\Delta_{Z_i}<.875$, and that the range of $\Delta_{Z_i}$ is smaller in this plot than it is in the $d=3$ central charge plots. These bounds were computed using $\ell_\text{max}=25$ and $\Lambda=19$. }
 \end{center}
 \end{figure}

Let us now present our bounds on the dimension $\Delta'_{Ss,0}$ of the lowest-lying $O(N)$-singlet scalar in the $Z_i\times\bar{Z}_j$ OPE. In Figure \ref{fig:3dbin} we show upper bounds on $\Delta'_{Ss,0}$ as a function of $\Delta_{Z_i}$ in $d=3$ and $d=3.5$. In both cases there are kinks in the bounds near the localization values of $\Delta_{Z_i}$. In this plot, the bound on $\Delta'_{Ss,0}$ at the localization values of $\Delta_{Z_i}$ is far above the estimated perturbative value for $\Delta'_{Ss,0}$ computed as a $\text{Pad\'e}_{[1,2]}$  extrapolation of the $4-\epsilon$ expansion results (black asterisks). 

 \begin{figure}[t]
\begin{center}
 \includegraphics[width = 0.8 \textwidth]{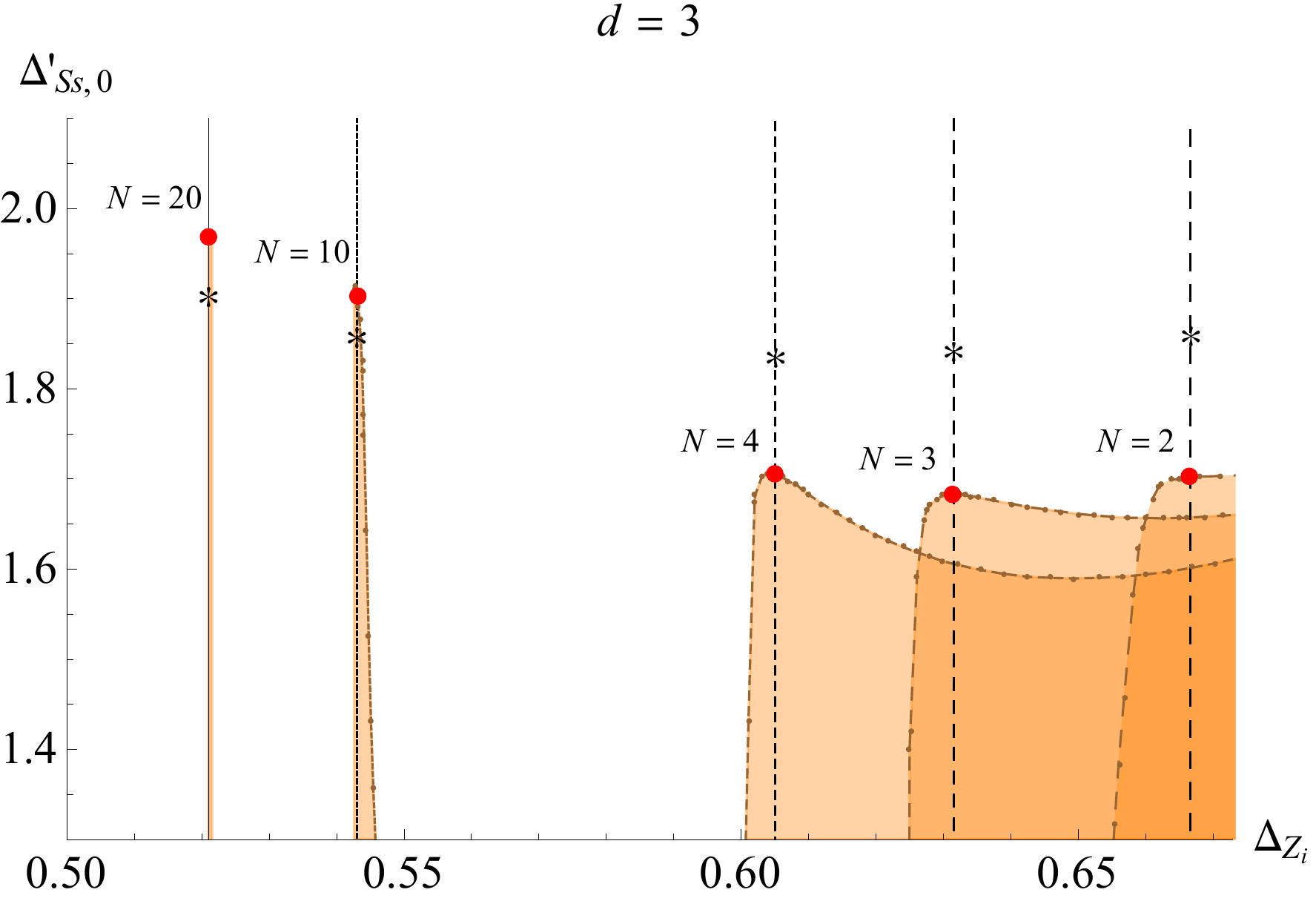}
\caption{\label{fig:mALL}Upper bound on the scaling dimension $\Delta'_{Ss, 0}$ of the lowest-lying $O(N)$-singlet scalar in the $Z_i\times\bar{Z}_j$ OPE, as a function of $\Delta_{Z_i}$ in dimensions $d=3$ and $d=3.5$, for $N = 2, 3, 4, 10, 20$, with the values of the central charges (Table \ref{val3}) imposed. Different shadings of orange denote the allowed regions for each $N$. The \textit{black} vertical lines denote localization values of $\Delta_{Z_i}$ for each $N$ (Table \ref{val3}) . The red dots denote the bounds at the localization values of $\Delta_{Z_i}$, as in Figure \ref{fig:3dbin}. The asterisks indicate the Pad\'e approximation of the 3-loop $\epsilon$-expansion values of $\Delta'_{Ss, 0}$ for each $N$. Note that the range of $\Delta_{Z_i}$ is smaller in this plot than it is in the $d=3$ central charge plots. These bounds were computed using $\ell_\text{max}=25$ and $\Lambda=19$. }
 \end{center}
 \end{figure}

The results improve dramatically after imposing the values of the central charges given in Table \ref{val3}.\footnote{We found no change in our results when imposing the localization value of $c_T$ in $d=3$ in addition to the other central charges.} 
As shown in Figure \ref{fig:mALL}, this extra input creates a sharp peak in the bounds around the localization values of $\Delta_{Z_i}$, essentially fixing them very precisely at larger values of $N$. In $d=3.5$, the bootstrap bounds with central charges imposed are within a couple of percent to the values of $\Delta'_{Ss,0}$ computed from the $\text{Pad\'e}_{[1,2]}(\epsilon)$ approximant of the  $\epsilon$-expansion for all the values of $N$ that we considered. In $d=3$, however, these bounds exclude the estimate from the 3-loop $\epsilon$-expansion when $N=2,3,4$. As mentioned in Section \ref{EPSILON}, 4-loop results for $\Delta'_{Ss,0}$ are also available in the literature, though we have found that different Pad\'e resummations of those give variable predictions, some of which are still excluded by our numerical bounds. While it is conceivable that a more sophisticated resummation method will improve the approximations based on the $\epsilon$-expansion, it is in general difficult to know a priori when these approximations are reliable. Predictions based on the numerical bootstrap, however, are rigorous and can serve as a litmus test for the validity of approximations.

\subsection{Bounds on $\Delta_{Z_i}$ Assuming $\lambda_{Z^2}=0$}
\label{chiralBounds}
 
   \begin{figure}[t]
\begin{center}
 \includegraphics[width = 0.85\textwidth]{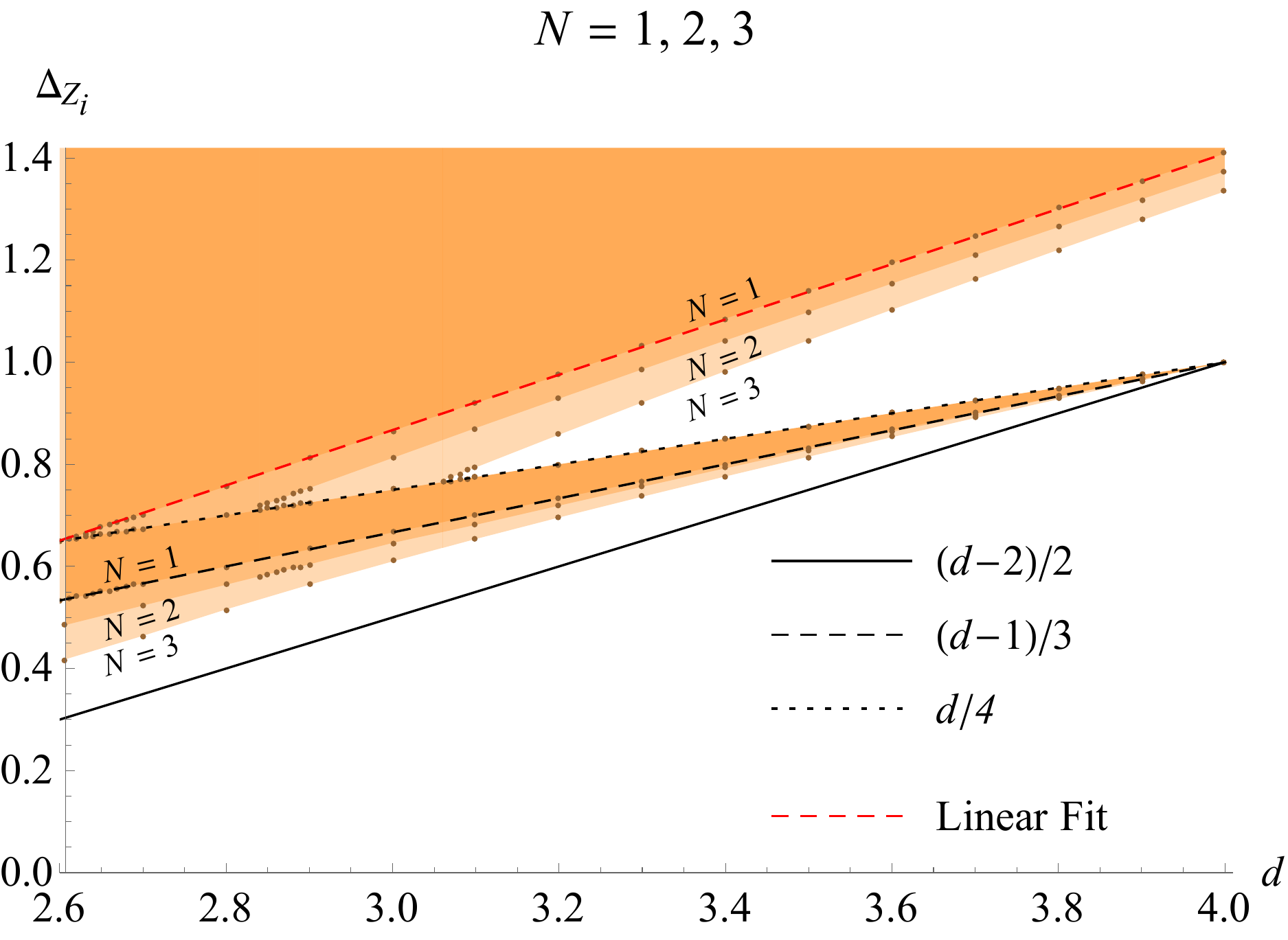}
 \caption{Allowed region for the scaling dimension $\Delta_{Z_i}$ of the $O(N)$-fundamental chiral primary $Z_i$ as a function of the dimension $d$ for $N=1,2,3$ (top to bottom), in the absence of a chiral primary in the $O(N)$-singlet channel of the $Z_i\times Z_j$ OPE\@. Different shadings of orange denote the allowed regions for each $N$. These bounds were computed using $\ell_\text{max}=25$ and $\Lambda=19$. \label{fig:delZplot}}
 \end{center}
 \end{figure}

The bounds on central charges presented in the previous subsections all showed a disallowed region for low enough values of $\Delta_{Z_i}$, and thereby provide lower bounds on it. These bounds on $\Delta_{Z_i}$ arise from the extra assumptions imposed on the operator spectrum due to the superpotential \eqref{Wtilde}, as discussed in Section \ref{BOOTSTRAP}. In particular, the assumption that the chiral operator $\sum_i Z_i^2$ is not a primary excludes the free theory, so a lower bound on $\Delta_{Z_i}$ strictly above unitarity is possible.  

In Figure \ref{fig:delZplot} the allowed range of $\Delta_{Z_i}$ is plotted against the dimension $d$ for $N=1,2,3$, assuming that the coefficient $\lambda_{Z_i^2}$ of the operator $\sum_i Z_i^2$ in the $Z_i\times Z_j$ OPE vanishes. For $N>1$ this assumption is included in the bootstrap setup described in Section \ref{BOOTSTRAP} and used throughout this study,\footnote{Inclusion of an operator with dimension $d-2\Delta_{Z_i}$ in the $Tt$ sector did not change results of this section.} while for $N=1$ we use the same bootstrap setup as in \cite{Bobev:2015jxa} supplemented with the condition $\lambda_{Z^2}=0$. 

For $N=1$ and $d \gtrsim 2.6$, there are two disconnected intervals of allowed values of $\Delta_{Z}$. The three boundaries of these intervals can be precisely identified with the three kinks observed in \cite{Bobev:2015jxa}. The bottom boundary of the bottom allowed region corresponds to the theory with superpotential $W=X^3$, which has $\Delta_{X}=(d-1)/3$. The top boundary of the bottom allowed region corresponds to the ``second kink'' observed in \cite{Bobev:2015jxa} appearing precisely when $\Delta_{Z}=d/4$.\footnote{Such gaps at $\Delta_{Z_i}=d/4$ appear for all values of $N$.}  As seen in Table \ref{UnitarityTable}, the position of the second kink is kinematically special, since the existence of a $\Delta= d-2\Delta_Z$ operator is consistent with unitarity only for $\Delta_{Z}\leq d/4$.
Finally, the bottom boundary of the top allowed region corresponds to the ``third kink'' discussed in \cite{Bobev:2015jxa}. The same kink was first observed in $d=4$ in \cite{Poland:2011ey} and it was recently explored further in \cite{poland2015exploring}. Curiously, as we emphasize with a linear fit in figure \ref{fig:delZplot}, the location of this third kink changes linearly as the dimension $d$ is increased. 

In a given dimension $d$ we always find two disconnected allowed regions for $\Delta_{Z_i}$ whenever $N < N_{\text{crit.}}(d)$, where the value of $N_{\text{crit.}}(d)$ increases with $d$. For $N > N_{\text{crit.}}(d)$ the second and third kinks do not exist and there is only one allowed region. As $d\to 4$, the bottom wedge  of the allowed region shrinks to zero size and $N_{\text{crit.}}(d)\to\infty$; i.e., in $d=4$ the first and second kink disappear, leaving us with one connected allowed region starting with what was the third kink in $d<4$, which now exists for all $N$. For $N=2,\, 3$, the bottom boundary of the bottom allowed region in figure \ref{fig:delZplot} is no longer linear and rather coincides with the localization values predicted for the models \eqref{Wtilde}. 

\begin{figure}[t]
	\begin{center}
	 \includegraphics[width = 0.85\textwidth]{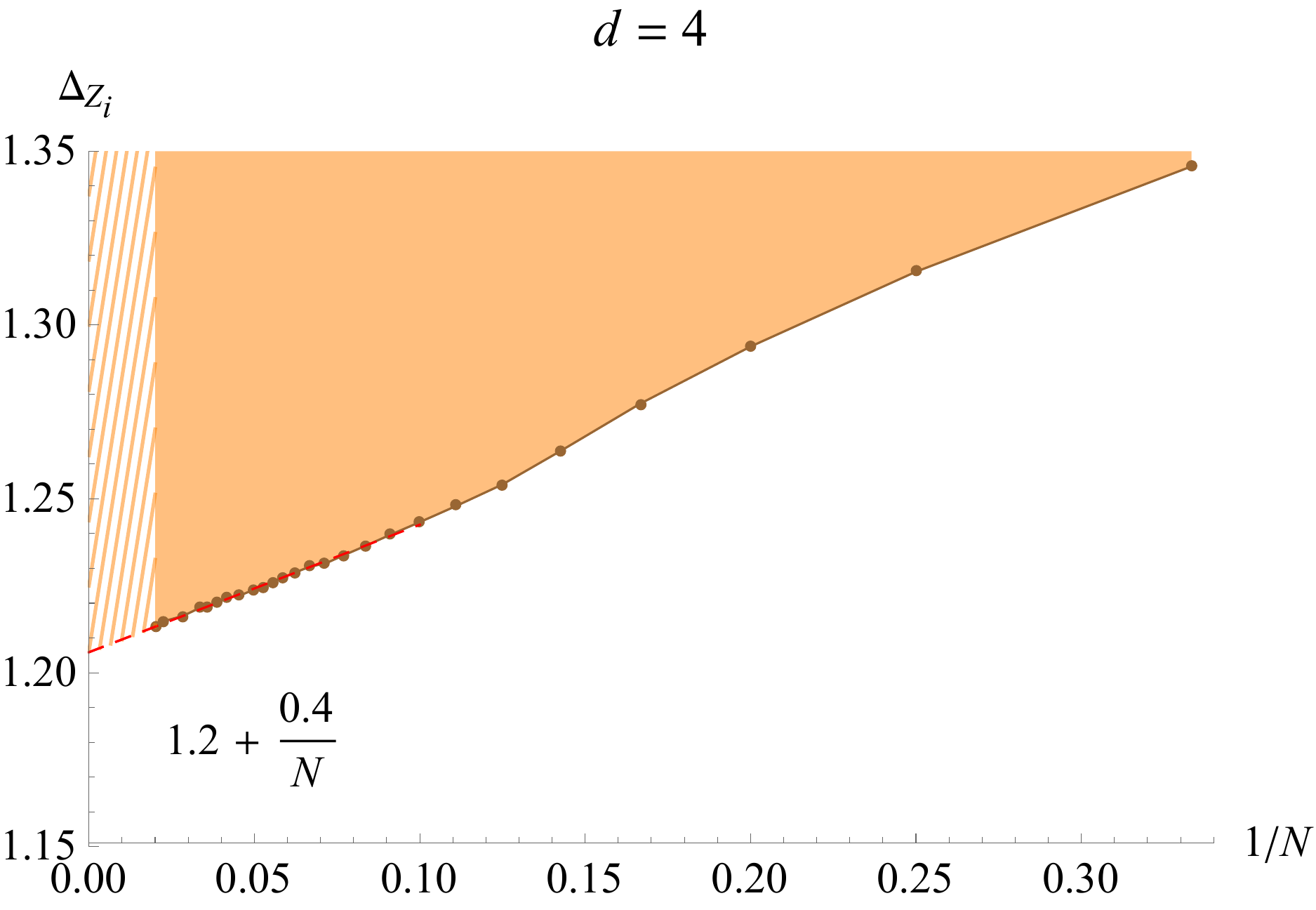}
	 \caption{Lower bound on the dimension $\Delta_{Z_i}$ of an $O(N)$-fundamental chiral primary $Z_i$ as a function of $1/N$ in $d=4$ for $3\leq N \leq50$, when $\sum_i Z_i^2$ is removed from the $Z_i\times Z_j$ OPE\@. The red dotted line denotes the linear extrapolation to $N\to\infty$. Solid orange denotes the allowed region, while textured orange denotes the allowed region given by the linear extrapolation of the lower bound.   \label{fig:4d}}
	 \end{center}
	 \end{figure}

In Figure \ref{fig:4d} we plot the position of the $d=4$ kink in terms of $1/N$ for $3\leq N\leq50$, so that we may determine the large $N$ behavior of the (as of yet) unknown SCFT, if any, that may correspond to this feature. For larger values of $N$, the numerical convergence of our results decreases drastically and therefore we rather extrapolate from lower values of $N$. The extrapolation from the roughly linear region $15\leq N\leq50$ to $N\to\infty$, yields an estimate for the scaling dimension of the chiral operator 
\es{4dlargeN}{
\Delta_{4d}=1.2+\frac{0.4}{N}+O(1/N^2)\,.
}
Unlike the other numerics in this study, for these numerics we used the improved bootstrap parameters $\ell_\text{max}=35$ and $\Lambda=27$, for which our bounds have a numerical uncertainty of $\Delta_{Z_i}=.001$. The linear extrapolation may have greater uncertainty, so we conservatively show our results to order $O(10^{-2})$.

\section*{Acknowledgments}

We thank Simone Giombi and Igor Klebanov for useful discussions. SSP and RY would like to thank the organizers of the Simons Summer Workshop in Mathematics and
Physics 2015, where some of this work was done, for their kind hospitality. This work was supported in part by the US NSF under Grant No.~PHY-1418069 (SMC and SSP) and PHY-1314198 (RY)\@.

\bibliographystyle{ssg}
\bibliography{Biblio}

\end{document}